\documentclass[a4paper,11pt]{article}

\usepackage{jheppub}
\usepackage{graphicx}
\usepackage{amsmath}
\usepackage{mathrsfs}
\usepackage{xspace}

\newcommand{\eq}[1]{eq.~\eqref{eq:#1}}
\newcommand{\eqs}[2]{eqs.~\eqref{eq:#1} and \eqref{eq:#2}}
\renewcommand{\sec}[1]{sec.~\ref{sec:#1}}
\newcommand{\secs}[2]{secs.~\ref{sec:#1} and \ref{sec:#2}}
\newcommand{\app}[1]{app.~\ref{app:#1}} 
\newcommand{\fig}[1]{fig.~\ref{fig:#1}}

\newcommand{\ord}[1]{{\mathcal O}(#1)}

\newcommand{\nn}{\nonumber}

\newcommand{\df}{\mathrm{d}}

\newcommand{\sdt}{\!\cdot\!}

\newcommand{\al}{\alpha}

\newcommand{\ga}{\gamma}
\newcommand{\Ga}{\Gamma}
\newcommand{\de}{\delta}

\newcommand{\eps}{\epsilon}

\newcommand{\Tau}{\mathcal{T}}

\newcommand{\bT}{\mathbf{T}}
\newcommand{\bn}{{\bar{n}}}

\newcommand{\IR}{\mathrm{IR}}
\newcommand{\UV}{\mathrm{UV}}

\newcommand{\bound}{\mathrm{bound}}

\newcommand{\one}{{(1)}}

\newcommand{\SCETa}{\ensuremath{{\rm SCET}_{\rm I}}\xspace}

\allowdisplaybreaks[2]

\title{Calculating Soft Radiation at One Loop}

\author[a,b]{Tomas Kasemets,}
\author[a,c]{Wouter J.~Waalewijn,}
\author[a]{Lisa Zeune}

\affiliation[a]{Nikhef, Theory Group, Science Park 105, 1098 XG, Amsterdam, The Netherlands}
\affiliation[b]{Department of Physics and Astronomy, VU University, De Boelelaan 1081, 1081 HV, Amsterdam, the Netherlands}
\affiliation[c]{ITFA, University of Amsterdam, Science Park 904, 1018 XE, Amsterdam, The Netherlands}

\emailAdd{kasemets@nikhef.nl}
\emailAdd{w.j.waalewijn@uva.nl}
\emailAdd{lisa.zeune@nikhef.nl}

\preprint{\vbox{
\hbox{NIKHEF 2015-043}}}

\abstract{
We present an efficient way to calculate the effect of soft QCD radiation at one loop, which is needed for predictions at next-to-next-to-leading logarithmic accuracy. We use rapidity coordinates and isolate the divergences in the integrand. By performing manipulations with cumulative variables, we avoid complications from plus distributions. We address rapidity divergences, divergences with an azimuthal dependence, complicated jet boundaries and multi-differential measurements. The process and measurements can be easily adjusted, as we demonstrate by  reproducing many existing soft functions. The results for a general LHC process with multiple Wilson lines are obtained by treating Wilson lines that are not back-to-back using a boost. We also obtain, for the first time, the $N$-jettiness soft function for generic jet angularities, and the collinear-soft function for the measurement of two angularities.}

\begin{document}

\maketitle

\section{Introduction}
\label{sec:intro}

LHC analyses involve restrictions on QCD radiation to increase their sensitivity. Restrictions can be imposed directly by e.g.~requiring a specific number of jets, or indirectly through e.g.~the transverse momentum of a Higgs boson. This leads to large logarithms in the cross section, requiring resummation to obtain reliable predictions. The origin of these large logarithms is the enhancement of collinear and soft radiation, which are treated as dynamic degrees of freedom in Soft-Collinear Effective Theory (SCET)~\cite{Bauer:2000ew, Bauer:2000yr, Bauer:2001ct, Bauer:2001yt}. SCET is an effective theory of QCD that achieves resummation through the factorization of hard, collinear and soft radiation at the level of the Lagrangian. 

In this paper we focus on soft radiation, which is encoded in the soft function in SCET.
The soft function is (schematically) defined as the matrix element 
\begin{align} \label{eq:soft_def}
\widehat S(m,\mu)
=\Big\langle 0 \Big |  {\rm\bar T}\Big[\prod\limits_i \widehat{Y}^{ \dagger}_i \Big]\, \delta(m - \hat m)\  {\rm T}\Big[\prod\limits_i \widehat{Y}_i \Big] \Big |  0 \Big \rangle\,, 
\end{align}
where $Y_i$ is a soft Wilson lines along the light-like direction of, and in the color representation of the $i$-th colored parton participating in the hard scattering. The T $(\rm \bar T$) denote (anti-) time ordering and the delta function encodes the measurement $m$ through the operator $\hat m$. 

We will present an efficient approach to calculate the one-loop soft function, which is an essential ingredient for resummation at next-to-next-to-leading logarithmic accuracy. We will not restrict ourselves to a specific process or measurement and demonstrate the versatility of our method by reproducing the one-loop soft function for thrust~\cite{Schwartz:2007ib,Fleming:2007xt}, angularities~\cite{Hornig:2009vb, Larkoski:2014uqa}, transverse momentum~\cite{Chiu:2012ir} and transverse thrust~\cite{Becher:2015gsa}. Results for the double differential measurement of two angularities and of transverse momentum and beam thrust are also reproduced~\cite{Larkoski:2014tva,Procura:2014cba}. These require an extension of SCET, called SCET${}_+$~\cite{Bauer:2011uc, Procura:2014cba,Larkoski:2015zka,jethierarchies}, with additional collinear-soft modes. The collinear-soft function is again a matrix element of eikonal Wilson lines and can be calculated in the same way.
We present for the first time the calculation of $N$-jettiness with generic jet angularities and the collinear-soft function for the double angularity measurement.

Our approach involves a combination of several tricks: We use the coordinates $k_T$, $y$ and $\phi$ that make the symmetries of the soft matrix element manifest. By isolating divergences at the integrand level, the integrals are simplified. In particular, the integral for the finite terms can directly be written down and evaluated numerically, if desired.
We work with the cumulative soft function, as this involves simple manipulations with logarithms rather than plus distributions. The soft function can be obtained by differentiating the final result. The $N$-jet soft function is given by the sum over emissions between all pairs of Wilson lines at one loop. We employ a boost to make the Wilson lines back-to-back, allowing us to recycle our dijet results. An extension of the hemisphere decomposition of ref.~\cite{Jouttenus:2011wh} is needed to handle more complicated boundaries between jets. Our approach is very general, as we also treat rapidity divergences and divergences with an azimuthal-angle dependence. In the latter case we find it convenient to use a version of dimensional regularization that has no $\eps$-dependence associated with the azimuthal angle, and show that this is consistent.

The calculation also provides insight into the structure of the soft function at one loop. For example, rapidity divergences are simply the divergences as the rapidity of the soft gluon goes to infinity. The divergent structure near the Wilson lines is dominated by the asymptotic behavior of the measurement. On the other hand, the divergences away from the Wilson lines depend on the area in $(y,\phi)$-space on which the measurement is defined, but are independent of the measurement itself.

The outline of the paper is as follows: In \sec{framework} we present the setup of our calculation. We discuss detailed examples for dijets observables in \sec{dijet},  generalized $N$-jettiness in \sec{1jettiness}, and double differential measurements in \sec{mdiff}. The conclusions are in \sec{conclusions}, and additional details related to the Becher-Bell rapidity regulator, the calculation of the jet function for transverse thrust and the results for thrust-like N-jettiness are relegated to the appendices.

\section{Calculational framework}
\label{sec:framework}

In this section we develop our calculational framework. We first describe the measurements we consider and the rapidity coordinates we use to express them. In \sec{dijetframework} we consider the one-loop soft function for (back-to-back) dijets and present our master formula in \eq{master_eta}. We extend this to $N$-jet production in \sec{not_back_to_back}, boosting to frames where Wilson lines are  back-to-back. Multi-differential measurements are discussed in \sec{mdiffframework}.

\subsection{Measurement function and rapidity coordinates}

For two back-to-back jets, the corresponding soft radiation is emitted from back-to-back Wilson lines. Its boost invariance is made manifest by describing the emitted gluon using its transverse momentum $k_T$, rapidity $y$ and azimuthal angle $\phi$. We will denote the contribution of this soft gluon to a measurement by a function $f(k_T,y,\phi)$, and require that the measurement is additive when there are multiple emissions (avoiding clustering effects from jet algorithms, see e.g.~\cite{Banfi:2012yh}). 

Collinear safety implies that for two partons in the collinear limit,
$f(k_{1T},y,\phi) + f(k_{2T},y,\phi) = f(k_{1T}+k_{2T},y,\phi)$. Consequently, 
\begin{align}
f(k_T,y,\phi) = k_T f(y,\phi)
\,.\end{align}
For a parton in the presence of a soft gluon $f(y',\phi') = f(y,\phi) + \ord{k_T^\text{soft}/k_T}$. In the soft limit $k_T^\text{soft} \to 0$ the deflection $y' - y$ and $\phi'-\phi$ due to the soft gluon go to zero, from which we conclude that IR safety imposes that $f(y,\phi)$ is continuous.
We will further assume that $f(y,\phi)\geq 0$, such that the measurement restricts the QCD radiation.
To rewrite measurements in these coordinates, we use
\begin{align}
  k^\mu &= k_T (\cosh y, \cos \phi, \sin \phi, \sinh y)
  \,, \quad
  k^- = k_T\, e^y
  \,, \quad 
  k^+ = k_T\, e^{-y} \,.
   \end{align}
Here $k^{\mp} = k^0 \pm k^3$ denote light-cone coordinates along the back-to-back jets, aligned with the $z$-axis.

\subsection{Dijets}
\label{sec:dijetframework}

We find it convenient to calculate the cumulative distribution for the soft function in terms of the measurement $m$ to avoid dealing with plus distributions
\begin{align}
  \de[m- k_T f(y,\phi)] &\to  \theta[m- k_T f(y,\phi)]
  \,, &
  \frac{1}{\mu}\, \frac{1}{(m/\mu)_+} &\to \theta(m)\,\ln \frac{m}{\mu}
  \,, & \nn \\
    \de(m) &\to \theta(m)
  \,, &
  \frac{1}{\mu}\, \Big(\frac{\ln (m/\mu)}{m/\mu}\Big)_+ &\to \frac12 \theta(m)\,\ln^2 \frac{m}{\mu}
\,,
&\ \text{etc.}
\end{align}
This simplifies intermediate steps, especially for multi-differential measurements. Of course, the distribution follows from differentiating the final expression with respect to $m$ and typically does contain plus distributions.

The calculation of the soft function will be carried out using dimensional regularization for both the UV and IR divergences, causing the virtual contributions to vanish ($1/\eps_\UV - 1/\eps_\IR=0$) and avoiding complications~\cite{Collins:1999dz, Manohar:2006nz, Lee:2006nr, Idilbi:2007ff} from the overlap with collinear radiation.
The real emission diagrams with the gluon attaching to Wilson lines 1 and 2 yield at this order (see also app.~C of ref.~\cite{Hoang:2014wka}) 
\begin{align} \label{eq:start}
S_{12}^\one(m,\mu)
& = -\frac{\al_s}{\pi^2}\, \bT_1 \sdt \bT_2\, \frac{\big(e^{\ga_E} \mu^2 \big)^\eps}{\Ga(1-\eps)}\, \nu^\eta
        \int_0^\infty\! \frac{\df k_T}{k_T^{1+\eta+2\eps}} \int_{-\infty}^\infty  \frac{\df y}{|2 \sinh y|^{\eta}} \int_0^{2\pi}\! \df \phi\, 
   \theta[m - k_T f(y,\phi)]
 \nn \\ &
 = \frac{\al_s}{\pi^2}\, \bT_1 \sdt \bT_2\, \frac{e^{\eps \ga_E}}{(\eta+2\eps) \Ga(1-\eps)}\,\frac{\nu^\eta \mu^{2\eps}}{m^{\eta+2\eps}}
     \int_{-\infty}^\infty\!\df y \int_0^{2\pi}\! \df \phi\, \frac{\theta[f(y,\phi)] f(y,\phi)^{2\eps}}{|2 \sinh y|^{\eta}}
\,.\end{align}
Here $\bT_1$ and $\bT_2$ denote the color charge of the emitted gluon in the representation of Wilson lines 1 and 2, respectively (in the notation of refs.~\cite{Catani:1996jh,Catani:1996vz}). If there are only two Wilson lines,
$\bT_1 \cdot \bT_2 = - C_F$  for a quark-anti-quark and $-C_A$ for two gluons. From \eq{start} it is clear that the soft radiation is uniformly emitted in $y$ and $\phi$. Thus if $f(y,\phi)$ goes to a constant for $y \to \pm \infty$, the $y$ integral diverges. We control these rapidity divergences in the soft function using the $\eta$ regulator of refs.~\cite{Chiu:2011qc,Chiu:2012ir}. Other regulators are possible~\cite{Collins:1981uk,Dixon:2008gr,Chiu:2009yx,Collins:2011zzd,Becher:2011dz,Echevarria:2015usa,Echevarria:2015byo}, and the expression corresponding to \eq{start} for ref.~\cite{Becher:2011dz} is given in \app{becher}. Note that at this order there is no distinction between outgoing and incoming Wilson lines, which is known to extend to two loops in certain cases~\cite{Kang:2015moa}.

We introduce a function $f_\infty(y,\phi)$ that captures the behavior of the measurement as $y \to \pm \infty$, such that $\ln (f/f_\infty)$ is integrable. In practice, $f_\infty$ can be obtained by expanding $\ln f$ around $1/y = 0$. 
This allows us to already isolate the divergent behavior at the integrand level, resulting in 
\begin{align}\label{eq:master_eta}
S_{12}^\one(m,\mu)
 &= \frac{\al_s}{2\pi^2}\, \bT_1 \sdt \bT_2\, 
  \int_{-\infty}^\infty\!\df y \int_0^{2\pi}\! \df \phi\, \theta[f(y,\phi)]\, f_\infty(y,\phi)^{2\eps} e^{-\eta |y|}
 \nn \\ & \quad \times
 \Big[\frac{1}{\eps} + 2 \ln \frac{\mu\,f(y,\phi)}{m\,f_\infty(y,\phi)} + 2\eps \Big(\ln^2 \frac{\mu}{m} - \frac{\pi^2}{24}\Big)\Big]
 \Big[1 + \eta \Big(-\frac{1}{2\eps} + \ln \frac{\nu}{m}\Big)\Big]
\,.\end{align}
The UV divergences are fixed by $f_\infty$ and the original measurement $f$ only enters in the finite terms through $\ln (f/f_\infty)$. At this order, only the asymptotic behavior of the rapidity regulator enters, which is characterized by the (simpler) factor $e^{-\eta|y|}$. 

An exception is when $f$ vanishes in regions of phase-space (see \eq{outside}).
In these cases it is convenient to separate $f$ into the measurement $f^M>0$ and the theta function $f^R$ defining the integration region.  Eq.~\eqref{eq:master_eta} now reads 
\begin{align}
S_{12}^\one(m,\mu)
 &= \frac{\al_s}{2\pi^2}\, \bT_1 \sdt \bT_2\, 
  \int_{-\infty}^\infty\!\df y \int_0^{2\pi}\! \df \phi\, f^R(y,\phi)\, f_\infty(y,\phi)^{2\eps} e^{-\eta |y|}
 \nn \\ & \quad \times
 \Big[\frac{1}{\eps} + 2 \ln \frac{\mu\,f^M(y,\phi)}{m\,f_\infty(y,\phi)} + 2\eps \Big(\ln^2 \frac{\mu}{m} - \frac{\pi^2}{24}\Big)\Big]
 \Big[1 + \eta \Big(-\frac{1}{2\eps} + \ln \frac{\nu}{m}\Big)\Big]
\,.\end{align}
Now $f_\infty$ can be determined by only considering $f^M$ (but is irrelevant if the integration is cut off by $f^R$).
When the region  described by $f^R$ has a finite area $A$ in  $(y,\phi)$ space,
\begin{align}\label{eq:outside}
S_{12}^\one(m,\mu)
 &= \frac{\al_s}{2\pi^2}\, \bT_1 \sdt \bT_2\, 
  \int_{-\infty}^\infty\!\df y \int_0^{2\pi}\! \df \phi\, f^R(y,\phi)\,
 \Big[\frac{1}{\eps} + 2 \ln \frac{\mu\,f^M(y,\phi)}{m} \Big]
 \nn \\
 &= \frac{\al_s}{2\pi^2}\, \bT_1 \sdt \bT_2\, \frac{A}{\eps}
  + \ord{\eps^0} 
\,.\end{align}
Thus the divergence is independent of $f_\infty$ and just proportional to this area. This is the motivation behind the hemisphere decomposition used in \sec{1jettiness}.
We will present several applications for dijet observables in \sec{dijet}, demonstrating the efficiency of this approach.

\subsection{$N$ jets}
\label{sec:not_back_to_back}

To calculate the soft function for $N$ Wilson lines, we can simply sum over the contribution from each pair of Wilson lines using \eq{master_eta}. However, we need to take into account that the Wilson lines are no longer back-to-back, which we address by boosting to a frame where they are back-to-back. Using primed coordinates for the former and unprimed coordinates for the latter, a momentum $k^\mu$ transforms as
\begin{align}
  k'^\mu =  B(n_1',n_2') k^\mu 
\,.\end{align}
with
\begin{align} \label{eq:boost}
  B(n_1',n_2') = \begin{pmatrix} \ga & - \ga \vec \beta^{\,T} \\ -\ga \vec \beta \quad & \mathbf{1}+(\ga-1) \vec \beta \vec \beta^{\,T} / \vec \beta^{\,2} \end{pmatrix} 
\,, \qquad
 \vec \beta = - \frac{1}{2} (\hat n_1' + \hat n_2')
 \,, \qquad
 \ga = \frac{\sqrt{2}}{{\sqrt{n_1' \sdt n_2'}}}
\,,\end{align}
where $n_i'=(1,\hat n_i')$ ($i$=1,2) denote the directions of the Wilson lines.
The Wilson lines in the two frames simply transform into each other.
Applying the reverse boost to $n_1'$, $n_2'$, $\bar n_1'$ and $\bar n_2'$, 
\begin{align}  \label{eq:ns}
  \tilde n_1^\mu &= \big(\ga^{-1}, \tfrac12(\hat n_1'- \hat n_2')\big) 
  \,, &
  \tilde n_2^\mu &= \big(\ga^{-1}, \tfrac12(\hat n_2' - \hat n_1')\big)  
  \,, \nn \\
  \tilde {\bar n}_1^\mu &= -\tilde n_1^\mu + 2\ga (1,\vec \beta)
  \,, &
  \tilde {\bar n}_2^\mu &= -\tilde n_2^\mu + 2\ga (1,\vec \beta)
\,,\end{align}
so $\tilde n_1$ and $\tilde n_2$ are indeed back-to-back, though  $\tilde n_i$ and  $\tilde {\bar n}_i$ are not.
Because $\tilde n_i$ and $\tilde {\bar n}_i$ do not have the conventional $(1,\hat n)$ normalization, we wrote a tilde on the $n_i$ and $\bar n_i$, though this normalization is irrelevant for the Wilson lines. 
One can then convert the measurement between the two coordinates using Lorentz invariance of scalar products $n_i' \sdt k' = \tilde n_i \sdt k$. For $i=1,2$ this takes a particularly simple form
\begin{align}
n_1' \sdt k' = \ga^{-1} n_1 \sdt k\,, 
\quad
n_2' \sdt k' = \ga^{-1} n_2 \sdt k
\label{eq:scalarboost1}
\,.\end{align}

This approach requires modification in the presence of rapidity divergences, since the rapidity regulator is not boost invariant. For definiteness we first assume that only the Wilson line in the $n_1'$ direction requires rapidity regularization. For the exchange of a soft gluon between the Wilson lines in the $n_1'$ and $n_2'$ direction, the rapidity regulator is
\begin{align} \label{eq:rap_boost}
  \Big(\frac{\nu}{|\bn_1' \sdt k' - n_1' \sdt k'|}\Big)^\eta \stackrel{y \to \infty}{=}
  \Big(\frac{\nu}{2\ga k_T \sinh y} \Big)^\eta
\,.\end{align}
Although inserting \eq{ns} leads to complicated expressions, the asymptotic behavior is simple and is the only thing that matters at one-loop order. The Wilson line requiring the rapidity regularization is at $y = \infty$, so this is the only relevant limit ($y\to -\infty$ is regulated by dimensional regularization). Note that if instead the Wilson line $n_2$ required rapidity regularization, the final expression would still be the same. From this we conclude that we may  use our master formula by simply replacing $\nu \to \nu/\ga$. In the presence of additional Wilson lines requiring rapidity regularization, we in principle need a copy of the rapidity regulator for each direction\footnote{Even for Wilson lines in the $n_1$ and $\bn_1$ directions we can have separate regulators, since the rapidity divergences should be cancelled by the collinear radiation in the $n_1$ and $\bn_1$ direction, respectively~\cite{Echevarria:2012js}.} 
\begin{align}
  \prod_i  \Big(\frac{\nu_i}{|\bn_i' \sdt k' - n_i' \sdt k'|}\Big)^{\eta_i}
\end{align}
Ensuring that rapidity divergences corresponding to the $n_i'$ direction are controlled by $\eta_i$, by taking the other $\eta$'s to zero first, implies that \eq{rap_boost} still holds with $\nu \to \nu_i$ and $\eta \to \eta_i$. In particular, if at the end of the calculation we take all regulators equal, $\nu_i = \nu$ and $\eta_i = \eta$, we can simply do all calculations by replacing $\nu \to \nu/\ga$ in our master formula. 

We find our approach of boosting to back-to-back coordinates convenient as it allows us to recycle results, but it is not necessary. Direct calculations of soft functions with more than two Wilson lines and rapidity divergences have been carried out in e.g.~refs.~\cite{Liu:2013hba,Becher:2015gsa}.

\subsection{Multi-differential measurements}
\label{sec:mdiffframework}

We now consider multi-differential measurements, where large logarithms associated with additional scales arise and require resummation. The resummation can be achieved by an extension of SCET (SCET${}_+$) with additional collinear-soft and/or soft-collinear degrees of freedom~\cite{Bauer:2011uc, Procura:2014cba, Larkoski:2015zka, Larkoski:2015kga, Becher:2015hka, Neill:2015nya, Chien:2015cka, jethierarchies}.
Whereas the soft function defined in \eq{soft_def} depends on one measurement $m$, multi-differential measurements give rise to a soft function depending on multiple measurements
\begin{align} \label{eq:multimeas}
 \de(m - \hat m) \to \prod_i \de(m_i - \hat m_i)
\,.\end{align}
The collinear-soft radiation of SCET${}_+$ is described by a collinear-soft function, which is also a matrix element of (collinear-soft) Wilson lines. It can be calculated in the same manner, as we will show in \sec{mdiff}.

To incorporate the multiple measurements in the soft function, we extend the measurement to a vector
\begin{align}
 \vec m &= k_T \vec f(y,\phi)
\,.\end{align}
allowing us to write \eq{multimeas} for the cumulative soft function as
\begin{align}
 \prod_i \theta[m_i - k_T f_i(y,\phi)] &= \theta[\max_i \{f_i(y,\phi)\}] \prod_i \theta[m_i/f_i(y,\phi) - k_T]
\,.\end{align}
For a given $y$ and $\phi$ this is dominated by a single measurement $m_I$ that imposes the strongest constraint on $k_T$.
Regulating this dominant measurement for $y\to \pm \infty$ through $f_\infty$, we arrive at following expression for the soft function
\begin{align}
S_{12}^\one(\vec m,\mu)
 &= \frac{\al_s}{2\pi^2}\, \bT_1 \sdt \bT_2\, 
  \int_{-\infty}^\infty\!\df y \int_0^{2\pi}\! \df \phi\,  \theta[\max_i \{f_i(y,\phi)\}] \, f_\infty(y,\phi)^{2\eps} e^{-\eta |y|}
 \\ & \quad \times
 \Big[\frac{1}{\eps} + 2 \ln \frac{\mu\,f_I(y,\phi)}{m_I\,f_\infty(y,\phi)} + 2\eps \Big(\ln^2 \frac{\mu}{m_I} \!-\! \frac{\pi^2}{24}\Big)\Big]
 \Big[1 + \eta \Big(-\frac{1}{2\eps} + \ln \frac{\nu}{m_I}\Big)\Big]
\,. \nn\end{align}
We emphasize that the index $I$ denoting the dominant measurement generally depends on $y$ and $\phi$. The corresponding division of phase-space provides a natural way to do the integration.

In \sec{mdiff} we will apply this to several double-differential measurements. Specifically, the measurement of two angularities~\cite{Larkoski:2014tva} and the simultaneous measurement of transverse momentum and beam thrust~\cite{Jain:2011iu}.

\section{Dijet examples}
\label{sec:dijet}

We start by calculating the soft function for the thrust and angularity $e^+e^-$ event shapes in \secs{thrust}{ang}.
In \sec{pT} we determine the transverse momentum soft function for $pp\to Z+X$ (or $pp \to H+X$), which contains rapidity divergences. 
For transverse thrust in $e^+e^-$ collisions, discussed in \sec{tt}, the divergences depend on the azimuthal angle. We describe how to treat this in dimensional regularization without breaking the azimuthal symmetry.

\subsection{Thrust}
\label{sec:thrust}

Thrust is an $e^+e^-$ event shape defined through~\cite{Farhi:1977sg}
\begin{align}
\tau=1-T= \frac{1}{Q}\,\sum_i \text{min} \left\{ k_i^+,k_i^-\right\}
\end{align}
with $i$ running over the final-state particles and $Q$ being the total invariant mass. The contribution of soft radiation to the measurement $m=Q\tau$, corresponds to
\begin{align}
 f(y,\phi) = e^{-|y|}
\,.\end{align}
Since $f$ is particularly simple, we choose $f_\infty = f$, leading to
\begin{align}
  S^\one(m=Q\tau,\mu)
 &= -\frac{\al_s C_F}{\pi}\,
  \int_{-\infty}^\infty\!\df y\, e^{-2 \eps |y|}
 \Big[\frac{1}{\eps} + 2 \ln \frac{\mu}{m} + 2\eps \Big(\ln^2 \frac{\mu}{m} - \frac{\pi^2}{24}\Big)\Big]  
 \nn \\ &
 = -\frac{\al_s C_F}{\pi}\,
 \frac{1}{\eps}\,\Big[\frac{1}{\eps} + 2 \ln \frac{\mu}{m} + 2\eps \Big(\ln^2 \frac{\mu}{m} - \frac{\pi^2}{24}\Big)\Big]  
\,.\end{align}
Differentiating this leads to the result of refs.~\cite{Schwartz:2007ib,Fleming:2007xt}.

\subsection{Angularities}
\label{sec:ang}

The contribution of soft radiation to the measurement $m=Q \tau_a$ of the angularity~\cite{Berger:2003iw} 
\begin{align}
\tau_a & = \frac{1}{Q} \sum_i k_{i T}\, e^{-|y_i|(1-a)}
\end{align}
is described by
\begin{align}
   f(y,\phi) & = e^{-|y|(1-a)}
\,.\end{align}
This family of event shapes is infrared safe for $a<2$ and includes thrust ($a=0$) and broadening ($a=1$). For $a<2$ and $a\neq 1$, with $f_\infty = f$, we obtain~\cite{Hornig:2009vb, Larkoski:2014uqa}
\begin{align}
S^\one(m=Q \tau_a,\mu) &= -\frac{\al_s C_F}{\pi}\,
  \int_{-\infty}^\infty\!\df y\, e^{-2 \eps |y| (1-a)}
 \Big[\frac{1}{\eps} + 2 \ln \frac{\mu}{m} + 2\eps \Big(\ln^2 \frac{\mu}{m} - \frac{\pi^2}{24}\Big)\Big]  
 \nn \\ &
 = \frac{\al_sC_F}{\pi} \frac{1}{a-1} \frac{1}{\eps}
 \Big[\frac{1}{\eps} + 2 \ln \frac{\mu}{m} + 2\eps \Big(\ln^2 \frac{\mu}{m} - \frac{\pi^2}{24}\Big)\Big]
  \nn \\ &
 = \frac{\al_s C_F}{\pi} \frac{1}{a-1}   \Big[\ \frac{1}{\eps^2} +\frac{1}{\eps} \Big( \ln \frac{\mu^2}{Q^2} - 2 \ln \tau_a \Big) 
 \nn \\ &
  \quad \quad+ \frac{1}{2} \ln^2 \frac{\mu^2}{Q^2} -2  \ln \frac{\mu^2}{Q^2} \ln \tau_a + 2  \ln^2 \tau_a - \frac{\pi^2}{12}
 \Big]
\,.\end{align}
The case $a=1$ is equivalent with the transverse momentum measurement discussed next.

\subsection{Transverse momentum}
\label{sec:pT}

When the transverse momentum, $p_T$, of the soft radiation is measured, $f$ is trivial
\begin{align}
   f(y,\phi) &= f_\infty(y,\phi) = 1
\,. \end{align}
However, the calculation is slightly more complicated due to rapidity divergences arising from $y \to \pm \infty$ in the $y$ integration,
\begin{align}
  S^\one(m=p_T,\mu)
 &= -\frac{\al_s C_F}{\pi}\,
  \int_{-\infty}^\infty\!\df y \, e^{-\eta |y|}
 \Big[\frac{1}{\eps} + 2 \ln \frac{\mu}{m} + 2\eps \Big(\ln^2 \frac{\mu}{m} - \frac{\pi^2}{24}\Big)\Big]
 \nn \\ & \quad \times   
 \Big[1 + \eta \Big(-\frac{1}{2\eps} + \ln \frac{\nu}{m} \Big)\Big]
 \nn \\ &
 = -\frac{\al_s C_F}{\pi}\,
 \Big[\frac{1}{\eps} + 2 \ln \frac{\mu}{m} + 2\eps \Big(\ln^2 \frac{\mu}{m} - \frac{\pi^2}{24}\Big)\Big]  \Big(\frac{2}{\eta} - \frac{1}{\eps} + 2 \ln \frac{\nu}{m} \Big) 
\,.\end{align}
As the rapidity regulator $\eta$ should not regulate UV divergences, it must be taken to zero before $\eps$. 
For Wilson lines in the adjoint representation (gluons), $C_F \to C_A$. This agrees with the calculation in ref.~\cite{Chiu:2012ir}, when converting their $\vec p_T$ measurement.

\subsection{Transverse thrust in $e^+e^-$}
\label{sec:tt}

The transverse thrust event shape $T_\perp$~\cite{Banfi:2004nk}, is designed for hadron collisions, but has also been calculated for $e^+e^- \to 2$ jets \cite{Becher:2015gsa},
\begin{align} \label{eq:Tperp_def}
\tau_\perp = 1 - T_\perp  &= \max_{\vec{n}_\perp} \frac{\sum_i  |\vec{k}_{i\perp}| - |\vec{k}_{i\perp} \cdot \vec{n}_{\perp}|}{\sum_i |\vec{k}_{i\perp}|} 
\nn \\
&=  \frac{\sum_i |\vec{k}_{i\perp}| - |\vec{k}_{i\perp} \cdot \vec{n}_{\perp}|}{Q |\sin\theta|}
\,.\end{align}
Here the sum $i$ runs over the final-state particles and the transverse ($\perp$) is with respect to the electron-positron beam axis. In the second line, power suppressed contributions have been neglected in order to write it in terms of the angle $\theta$ between the beam and the thrust axis, and the transverse orientation of the thrust axis $\vec n_\perp$. The contribution to $\tau_\perp$ from one soft particle with momentum $k$ is thus described by
\begin{align}\label{eq:f_tau_perp}
f(y,\phi) & = \frac{1}{Q|\sin\theta|} \Big[ \sqrt{( \cos\phi \cos\theta + \sinh y \sin\theta )^2 + \sin^2\phi} - |\cos\phi \cos\theta + \sinh y \sin\theta | \Big]\, , \nn\\
f_\infty(y,\phi) & = \frac{\sin^2\phi}{Q \sin^2\theta} e^{-|y|} \,,
\end{align}
where  we have expressed $k_\perp$ and $n_\perp$ in \eq{Tperp_def} in terms of the variables $k_T$, $y$ and $\phi$ in the frame where the thrust axis is along the $\hat{z}$ direction.
Interestingly, the structure of the divergence as $y \to \pm \infty$ in \eq{f_tau_perp} has an azimuthal angle dependence. 
This results in
\begin{align}
S^\one(m=\tau_\perp,\mu)
 &= -\frac{\al_s C_F}{2\pi^2}
  \int_{-\infty}^\infty\!\df y \int_0^{2\pi}\! \df \phi\, 
   \\
  & \quad \times  \Big\{ f_\infty(y,\phi)^{2\eps}
 \Big[\frac{1}{\eps} + 2 \ln \frac{\mu}{m} + 2\eps \Big(\ln^2 \frac{\mu}{m} - \frac{\pi^2}{24}\Big)\Big] + 2\ln\frac{f(y,\phi)}{\,f_\infty(y,\phi)}  \Big\}
\nn \\ 
& = - \frac{\al_s C_F}{\pi} \Big[ \frac{1}{\eps^2} + \frac{2}{\eps} \ln \frac{\mu}{4mQ\sin^2\theta} + 2 \ln^2 \frac{\mu}{4mQ\sin^2\theta} + \frac{7\pi^2}{12} + A(\theta) \Big]
\,,\nn\end{align}
where the finite term $A(\theta)$ is given by
\begin{align}
A(\theta) & = \frac{1}{\pi} \int_{-\infty}^\infty\!\df y \int_0^{2\pi}\! \df \phi 
  \ln \frac{f(y,\phi)}{\,f_\infty(y,\phi)}
\,.\end{align}
We remind the reader that these $y$ and $\phi$ are defined in the frame where the thrust axis is along the $\hat{z}$ axis, while the {\it transverse} in transverse thrust means perpendicular to the beam axis. The results have been cross checked with ref.~\cite{Becher:2015gsa}, and agree once the different scheme for dimensional regularization is taken into account, as discussed in detail below.

The transverse part of the $d$-dimensional integration measure can be written
\begin{align}
\int\! \df^{2-2\epsilon} k_\perp = \frac{\Omega_{1-2\epsilon}}{2} \frac{1}{2}\int\! \df k_T^2 (k_T^2)^{-\epsilon} \int_0^{2\pi} \df\phi\, \bigl[\sin^2(\phi-\phi_0)\big]^{-\epsilon}
,\end{align}
where $\phi-\phi_0$ is the azimuthal angle between the momentum $k_\perp$ and an arbitrary reference axis. With the choice $\phi_0=0$ we obtain the integration measure used in ref.~\cite{Becher:2015gsa}.
We prefer to preserve the azimuthal symmetry, and integrate over the choice of this reference axis
\begin{align}
\int\! \df^{2-2\epsilon} k_\perp & =\frac{\Omega_{1-2\epsilon}}{2} \frac{1}{2}\int\! \df k_T^2 (k_T^2)^{-\epsilon}  \frac{1}{2\pi}\int_0^{2\pi}\! \df\phi_0  \int_0^{2\pi}\! \df\phi \big[\sin^2(\phi-\phi_0)\big]^{-\epsilon} 
\nn\\
 & = \frac{\Omega_{2-2\epsilon}}{2\pi} \frac{1}{2}\int\! \df k_T^2 (k_T^2)^{-\epsilon} \int_0^{2\pi}\! \df\phi.
\end{align}
When the measurement does not depend on $\phi$ the two ways gives the same results since
\begin{align}
\frac{\Omega_{1-2\epsilon}}{2} \int_0^{2\pi}\! \df\phi \bigl(\sin^2\phi\bigr)^{-\epsilon} = \frac{2\pi^{1-\epsilon}}{\Gamma(1-\epsilon)} = \Omega_{2-2\epsilon}. 
\end{align}
However, for transverse thrust, which does depend on the azimuthal angle, the two schemes give different results for the cumulative soft function. With  $f_\infty^{2\epsilon} \propto (\sin^2\phi)^{2\epsilon}$, the two measures lead to contributions to the cumulative soft function that are related through 
\begin{align}
-\frac{\alpha_s C_F}{2\pi^2}\,\frac{1}{\eps^2}\, \frac{\Omega_{2-2\epsilon}}{2\pi} \int_0^{2\pi}\! \df\phi\, (\sin^2\phi)^{2\epsilon} &= -\frac{\alpha_s C_F}{2\pi^2}\, \frac{1}{\eps^2}\,\frac{\Omega_{1-2\epsilon}}{2} \int_0^{2\pi}\! \df\phi\, (\sin^2\phi)^{\epsilon} - \frac{\alpha_s C_F}{\pi}\, \frac{2\pi^2}{3} + \ord{\epsilon}.
 \end{align}
The extra $\pi^2$ term in the finite part of the cumulative soft function is cancelled by corresponding terms in the two jet functions, calculated in \app{jetfunction}.

\section{$N$-jettiness with generic jet angularities}
\label{sec:1jettiness}

We extend the thrust-like $N$-jettiness definition~\cite{Stewart:2010tn,Jouttenus:2011wh}, by considering the measurement of a different angularity for each jet\footnote{Here we use the term `jets' to refer to both final-state and beam jets.} 
\begin{align} \label{eq:Tau_N}
\Tau_N=\sum_h \min_\ell \Big\{  \frac{2 \omega_\ell}{Q_\ell} (n_\ell' \sdt k_h')^{1-\alpha_\ell/2}(\bar{n}_\ell'\sdt k_h')^{\alpha_\ell/2} \Big\}
\equiv \sum_{\ell}\Tau^{\ell,\alpha_\ell}_N
\,,\end{align}
where $h$ runs over the hadronic final-state particles and $\ell$ over the jets in the event with (label) momenta 
\begin{align} \label{eq:q_l}
q_\ell' = \omega_\ell  n_\ell'  = \omega_\ell (1, \hat n_\ell' )\,.
\end{align}
The primed variables indicate that the momenta are defined in generic coordinates. We will later boost to (unprimed) coordinates where two of the Wilson lines are back to back, as discussed in \sec{not_back_to_back}. The $\omega_\ell$ in \eqs{Tau_N}{q_l} is considered a parameter which does not transform between frames (i.e.~no $\omega_\ell'$).
The minimization of \eq{Tau_N} assigns each particle to a jet region, and $\Tau^{\ell,\alpha_\ell}_N$ is the total contribution from jet region $\ell$.
The `standard' thrust-like $N$-jettiness definition is recovered if all $\alpha_\ell$ are zero, $\Tau^{\ell}_N \equiv \Tau^{\ell,\alpha_\ell=0}_N$. We show how our results reduce to the expressions in ref.~\cite{Jouttenus:2011wh} in \app{Njettiness}.
We will assume $\al_\ell \neq 1$ to avoid rapidity divergences.

The one-loop soft function is the sum over contributions from gluons exchanged between Wilson lines corresponding to the jets $i$ and $j$
\begin{align}
S^\one(m,\mu) = 
\sum_{i<j} S_{ij}^\one(m,\mu)
\,.\end{align}
To simplify the discussion we consider $1$-jettiness in $pp$ collisions (or equivalently $3$-jettiness in $e^+ e^-$ collisions). We label the three jets by $\ell=i,j,m$ to make the extension to $N$ jets straightforward.
The contribution of a soft gluon to $\Tau^{i,\alpha_i}_{1}$ is given by\footnote{To simplify the expressions for the measurement functions $f^M$, we already pull out a factor of the transverse momentum $k_T$ in the unprimed coordinates (where Wilson lines are back-to-back).}
\begin{align}
k_T \,k_{i}
\theta \big(k_{j}- k_{i}\big) \theta \big( k_{m}- k_{i}\big)\,,
\end{align}
and similarly for $\Tau^{j,\alpha_j}_{1}$ and $\Tau^{m,\alpha_m}_{1}$, where we introduced
\begin{align}
k_{\ell}=\frac{ 2 \omega_\ell}{Q_\ell k_T} (n_\ell'\sdt k')^{1-\alpha_\ell/2}(\bar{n}_\ell'\sdt k')^{\alpha_\ell/2} \,.
\end{align}

\begin{figure}[t]
  \centering
   \includegraphics[width=0.99\textwidth]{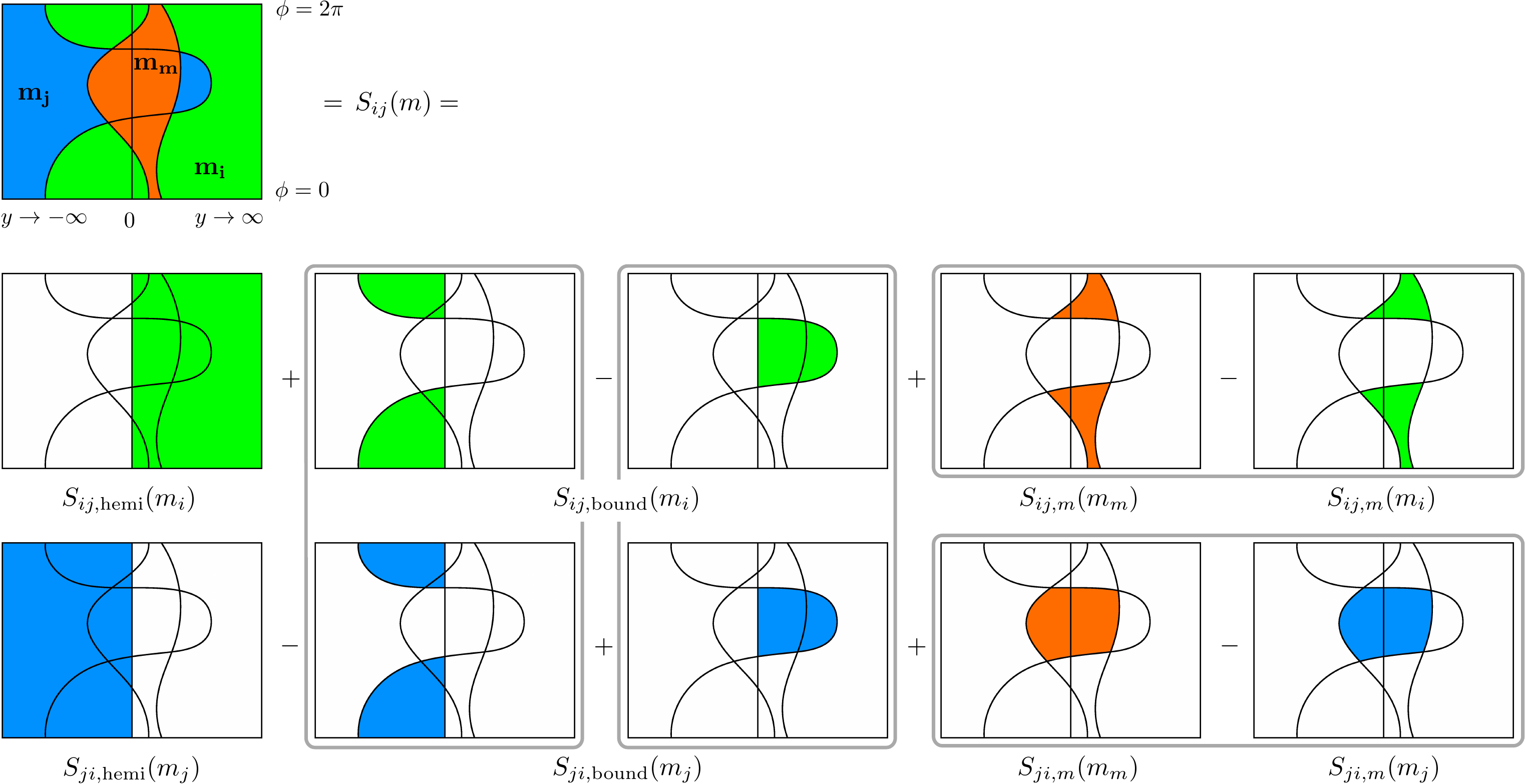}
   \caption{
Separation of the soft function, $S_{ij}$, with a gluon emitted between the $i$th and $j$th Wilson line, into
hemisphere, boundary and non-hemisphere contributions. The contributions surrounded by a gray box are together finite.}
   \label{fig:regions}
  \end{figure}
We extend the hemisphere decomposition~\cite{Jouttenus:2011wh} to handle the azimuthally dependent phase-space boundaries between regions arising from the more general $N$-jettiness measurement. This approach is discussed in detail in 
Ref.~\cite{softfunction}. The decomposition of $S_{ij}^\one$ into hemisphere, boundary and non-hemisphere contributions is depicted in \fig{regions} and will be discussed below.
The soft function involves three regions associated with the measurements: $\theta(k_j-k_i)\theta(k_m-k_i)$ for $m_i$, $\theta(k_i-k_j)\theta(k_m-k_j)$ for $m_j$ and $\theta(k_i-k_m)\theta(k_j-k_m)$ for $m_m$. With the purpose of making the analytical calculation of the divergent parts, as well as the extension to $N$ jets as easy as possible, we first allow the measurements of $m_i$ and $m_j$ to extend over the region of $m_m$. This is then compensated for by the non-hemisphere contributions $S_{ij,m}$ and $S_{ji,m}$ to the soft function. For a generic measurement, such as the one considered here, the separation between the regions for $m_i$ and $m_j$ is a non-trivial contour in ($y$, $\phi$)-space, but the divergencies of the soft function do not depend on the exact form of the contour. We therefore split the ($y$, $\phi$)-space into the two hemispheres  $y>0$ for $m_i$  and $y<0$ for $m_j$.  To compensate for the difference between cutting the phase space along $y=0$ compared to the original contour between $m_i$ and $m_j$, we introduce the boundary contribution $S_{ij,\bound}+S_{ji,\bound}$. Adding up these contributions,
\begin{align}
S_{ij}^\one(m=\{\Tau^{i,\alpha_i}_{1},\Tau^{j,\alpha_j}_{1},\Tau^{m,\alpha_m}_{1}\},\mu) &= S^\one_{ij,\rm{hemi}}(m_i=\Tau^{i,\alpha_i}_{1},\mu)  \,
 + S^\one_{ij,\bound}(m_i=\Tau^{i,\alpha_i}_{1}\},\mu) \nn \\&  \quad
 + S^\one_{ij,m}(m_m=\Tau^{m,\alpha_m}_{1},\mu)  \,
- S^\one_{ij,m}(m_i=\Tau^{i,\alpha_i}_{1},\mu)
 \nn \\&  \quad
 + (j  \leftrightarrow i)\,.
\end{align}
As we will see, the hemisphere contributions contain all divergencies, whereas the boundary and
non-hemisphere contributions are UV and IR finite.
When there are additional jets, the hemisphere and boundary contributions are of course the same, but there will be additional non-hemisphere contributions.

We now boost such that the Wilson lines $i$ and $j$ become back-to-back, allowing us to use \sec{not_back_to_back} to perform the calculation.
Using \eqs{ns}{scalarboost1}, this leads to the following expressions for the $k_\ell$ in the back-to-back frame
\begin{align}
& k_{i}= \frac{ 2 \omega_i}{Q_i} \gamma^{-1} e^{-y\,(1-\alpha_i/2)} \left(a e^{-y} + b e^y + c \cos(\phi-\phi_0)\right)^{\alpha_i/2}
\,, \nn \\
& k_{j}= \frac{ 2 \omega_j}{Q_j} \gamma^{-1} e^{y\,(1-\alpha_j/2)} \left(b e^{-y} + a e^y + c \cos(\phi-\phi_0) \right)^{\alpha_j/2}
\,, \nn \\
& k_{m}=\frac{2 \omega_m}{Q_m} \Big(\frac{1}{2} e^y (\tilde{n}^0_m-\tilde{n}^3_m) + \frac{1}{2} e^{-y} (\tilde{n}^0_m+\tilde{n}^3_m)-\tilde{n}^1_m \cos\phi-\tilde{n}^2_m \sin\phi \Big)^{1-\alpha_m/2} 
\, \nn \\
&\phantom{k_{m}\frac{2 \omega_m}{Q_m} }\times \Big(\frac{1}{2} e^y (\tilde{\bar{n}}^0_m-\tilde{\bar{n}}^3_m) + \frac{1}{2} e^{-y} (\tilde{\bar{n}}^0_m+\tilde{\bar{n}}^3_m)-\tilde{\bar{n}}^1_m \cos\phi-\tilde{\bar{n}}^2_m \sin\phi\Big)^{\alpha_m/2}
\,,\end{align}
with
\begin{align}
a = \gamma^2 - 1
\,, \qquad
b = \gamma^2
\,, \qquad
c = 2\gamma\sqrt{(\gamma^2-1)}
\,.\end{align}
Here we have explicitly chosen the $z$-axis through $\hat{z}=\tfrac12 \gamma (\hat{n}'_i-\hat{n}'_j)$. The azimuthal angle $\phi_0$ of the boost $- \vec{\beta}$ in \eq{boost} plays no role in the rest of the calculation. 

The measurement functions for the different jets are defined as
\begin{align}
f^M_i  = k_i \,, \qquad f^M_j  = k_j\,, \qquad  f^M_m  = k_m\,.
\end{align}
Starting with $S^\one_{ij,\rm{hemi}}$, we have 
\begin{align}
f^R_{\rm{hemi},i} =
\theta (y)
\,.\end{align}
The form of the measurement simplifies considerably in the limit of $y\rightarrow \infty$, in particular the dependence on the azimuthal angle vanishes
\begin{align}
f_{\infty,i} = 2 \frac{\omega_i}{Q_i} \gamma^{-1} b^{\alpha_i/2} e^{-(1-\alpha_i)y}\,.
\end{align}
The hemisphere contribution is now
\begin{align}\label{eq:hemii}
S^\one_{ij,\rm{hemi}}(m_i=\Tau^{i,\alpha_i}_{1},\mu)
&= \frac{\al_s}{\pi}\,\bT_i \sdt \bT_j
  \int_{-\infty}^\infty\!\df y\, 
  f^R_{\rm{hemi},i} \, f_{\infty,i}^{2\eps}
\nn \\ & \quad \times   
\Big[\frac{1}{\eps} + 2 \ln \frac{\mu}{m_i} + 2\eps \Big(\ln^2 \frac{\mu}{m_i} - \frac{\pi^2}{24}\Big)\Big]  + I_{\rm{hemi},i}
\nn \\ 
& = \frac{\al_s}{2\pi}\frac{1}{ (1- \alpha_i)}\,\bT_i \sdt \bT_j \Big( \frac{1}{\eps^2} +\frac{2}{\eps} \ln \frac{ B_i \mu}{m_i} + 2 \ln^2 \frac{ B_i \mu}{m_i} - \frac{\pi^2}{12} \Big) + I_{\rm{hemi},i} \,,
\end{align}
where
\begin{align}\label{eq:B}
B_i = 2\frac{\omega_i}{Q_i} \gamma^{-1}  b^{\alpha_i/2} \,,
\end{align}
and the remaining finite integral is
\begin{align}\label{eq:Ihemi}
I_{\rm{hemi},i} =   \frac{\al_s}{\pi^2}\,\bT_i \sdt \bT_j
  \int_{-\infty}^\infty\!\df y\, 
  \int_{0}^{2 \pi}\!\df \phi\,
  f^R_{\rm{hemi},i} \, \ln \frac{f^M_i}{f_{\infty,i}}\,.
\end{align}
The second hemisphere contribution $S^\one_{ji,\rm{hemi}}(m_j=\Tau^{j,\alpha_j}_{1},\mu)$, describing the region $ y<0$, 
is given by replacing $i\rightarrow j$ in the final line of \eq{hemii}.
 
Next we calculate the boundary contribution, shown in the second and third column of \fig{regions}. The integration over $y$ and $\phi$ is finite and we can use \eq{outside} to write
 \begin{align} \label{eq:non-hemiA}
 S^\one_{ij,\bound}(m_i,\mu)  &= \frac{\al_s}{2\pi^2}\,\bT_i \sdt \bT_j
  \int_{-\infty}^\infty\!\df y \int_0^{2\pi}\! \df \phi\,
  f^R_{ij,\bound}
\Big(\frac{1}{\eps} + 2 \ln \frac{\mu}{m_i} +  2\ln f^M_{i} \Big) \,,
 \end{align}
with 
\begin{align}
f^R_{ij,\bound} = \theta(- y)\theta(k_j-k_i) - \theta( y)\theta(k_i-k_j)\,,
\end{align}
The region for $S^\one_{ji,\bound}(m_j,\mu)$ is given by $f^R_{ji,\bound}=-f^R_{ij,\bound}$. Therefore, the area of the two contributions are equal but enter with different signs, such that the poles cancel in the combination. The total boundary contribution is thus
\begin{align}
 S^\one_{ij,\bound}(m_i,\mu) &+ S^\one_{ji,\bound}(m_j,\mu)  = \frac{\al_s}{2\pi^2}\,\bT_i \sdt \bT_j
  \int_{-\infty}^\infty\!\!\!\df y \int_0^{2\pi}\! \df \phi\, f^R_{ij,\bound} \Big( 2 \ln\frac{m_j}{m_i} + 2 \ln \frac{f^M_{i}}{f^M_{j}} \Big)\,.
\end{align}

The measurement functions for the non-hemisphere contributions,
$S^\one_{ij,m} (m_m,\mu)$ and $S^\one_{ij,m} (m_i,\mu)$, shown in the last two columns of \fig{regions},
are defined on the same region,
\begin{align}
f^R_{ij,m}=
\theta (k_{j} - k_{i})
\theta (k_{i} - k_{m})
\,.\end{align}
Application of \eq{outside} gives
\begin{align}
 S^\one_{ij,m}(m_{m}=\Tau^{m,\alpha_{m}}_{1},\mu) &= \frac{\al_s}{2\pi^2}\,\bT_i \sdt \bT_j
  \int_{-\infty}^\infty\!\df y \int_0^{2\pi}\! \df \phi\, f^R_{ij,m} 
\Big(\frac{1}{\eps} + 2 \ln \frac{\mu}{m_{m}} + 2 \ln f^M_{m} \Big) \,,
\end{align}
and similarly for $S^\one_{ij,m}(m_{i},\mu)$ with the replacement $m \rightarrow i$.
Subtracting the non-hemisphere $i$ contribution from the non-hemisphere $m$ contribution, the $1/\epsilon$ poles cancel and for 
the full non-hemisphere contribution we find
\begin{align}\label{eq:non-hemi}
S^\one_{ij,m}(m_m)  \,
- S^\one_{ij,m}(m_i) &=
\frac{\al_s}{\pi}\,\bT_i \sdt \bT_j \Big( \tilde{I}_0  \ln \frac{m_i}{m_m}  +  \tilde{I}_1 \Big)\,,
\end{align} 
 with
\begin{align}\label{eq:ints}
 \tilde{I}_0 &= \frac{1}{\pi} \int_{-\infty}^\infty\!\df y \int_0^{2\pi}\! \df \phi\, f^R_{m}  \,, \nn \\ 
 \tilde{I}_1 &= \frac{1}{\pi} \int_{-\infty}^\infty\!\df y \int_0^{2\pi}\! \df \phi\, f^R_{m}   \ln 
 \frac{f^M_{m}}{f^M_{i}} \,.
\end{align} 
Note that $\tilde I_0$ is simply the area of region $m$ with $k_i<k_j$. The result for the second non-hemisphere contribution $S^\one_{ji,m}(m_m)  \,
- S^\one_{ji,m}(m_j)$ is obtained by the replacement $i\leftrightarrow j$ in \eq{non-hemi} and \eq{ints}.
We show in \app{Njettiness} how for $\alpha_\ell=0$ these expressions reduce to those in  ref.~\cite{Jouttenus:2011wh}.

\section{Multi-differential measurements}
\label{sec:mdiff}

We present results for the soft function and the collinear-soft function for double differential measurements. In \sec{thrust_pT} we consider the simultaneous measurement of (beam) thrust and transverse momentum, and in \sec{tatb} the measurement of two angularities.

\subsection{Thrust and transverse momentum}
\label{sec:thrust_pT}

Following ref.~\cite{Procura:2014cba}, we combine the (beam) thrust and transverse momentum measurements of \secs{thrust}{pT}, which is described by
\begin{align}
 \vec f(y,\phi) &= (e^{-|y|},1)  
\,.\end{align}
In the asymptotic regime $y\to \pm \infty$ the transverse momentum measurement dominates,
\begin{align}
 f_\infty(y,\phi) = 1
\,, \end{align}
leading to~\cite{Procura:2014cba}
\begin{align} \label{eq:S_tau_pT}
S^\one(\vec m=(Q\tau,p_T),\mu)
 &= -\frac{\al_s C_F}{\pi}\, 
  \int_{-\infty}^\infty\!\df y\,  e^{-\eta |y|}
 \bigg[\frac{1}{\eps} + 2 \ln \frac{\mu}{\min(m_1 e^{|y|},m_2)} 
 + 2\eps \Big(\ln^2 \frac{\mu}{m_2}  - \frac{\pi^2}{24}\Big)\bigg]
 \nn \\ & \quad \times
 \Big[1 + \eta \Big(-\frac{1}{2\eps} + \ln \frac{\nu}{m_2}\Big)\Big]
\nn \\ &
= S^\one(m=p_T,\mu) -
\theta(m_2 - m_1)\,
 \frac{2\al_s C_F}{\pi}\, 
  \int_0^{\ln (m_2/m_1)}\!\df y\,  
 2 \ln \frac{m_2}{m_1 e^y} 
\nn \\ &
= S^\one(m=p_T,\mu) -
\theta(m_2 - m_1)\,
 \frac{2\al_s C_F}{\pi}\, \ln^2 \frac{m_2}{m_1}
\,.\end{align}
In the second step we first assumed that $\min(m_1 e^{|y|},m_2) = m_2$, leading to the transverse momentum soft function, and corrected for this through the second term.

The collinear-soft function for this double differential measurement is a matrix element of (collinear-soft) Wilson lines, and thus leads to the same amplitude as in  \eq{master_eta}. However, due to the collinear nature of this radiation, we use the measurement function for the hemisphere it goes into.\footnote{In the calculation one also integrates over the other hemisphere. This is corrected for through zero-bin subtractions~\cite{Manohar:2006nz} that remove the overlap with soft radiation, but vanish in pure dimensional regularization.}
For collinear-soft radiation going into the $y<0$ hemisphere, 
\begin{align}
 \vec f(y,\phi) &= (e^y,1) 
 \,, \quad
 f_\infty(y,\phi) = \theta(-y)+\theta(y)e^{y}
\,.\end{align}
We thus find
\begin{align}
{\mathscr S}^\one(\vec m=(p^-,p_T),\mu)
&= \frac12 S^\one(\vec m=(Q\tau,p_T),\mu)
 \nn \\ & \quad
-\frac{\al_s C_F}{\pi}\, 
  \int_{0}^\infty\!\df y\,  e^{2\eps y}
 \bigg[\frac{1}{\eps} + 2 \ln \frac{\mu}{m_1}
 + 2\eps \Big(\ln^2 \frac{\mu}{m_1}  - \frac{\pi^2}{24}\Big)\bigg]
 \nn \\ & \quad
-\theta(m_1 - m_2) \frac{\al_s C_F}{\pi}\, 
  \int_{0}^{\ln(m_1/m_2)}\!\df y\,  
   2 \ln \frac{m_1}{m_2 e^y}
\nn \\ &
= \frac12 S^\one(\vec m=(Q\tau,p_T),\mu)
-\frac{\al_s C_F}{\pi}\, 
\Big[- \frac{1}{2\eps^2} - \frac{1}{\eps} \ln \frac{\mu}{m_1} - \ln^2 \frac{\mu}{m_1} + \frac{\pi^2}{24}
\nn \\ & \quad
-\theta(m_1 - m_2) \ln^2 \frac{m_1}{m_2}\Big]
\,,\end{align}
exploiting that the measurement is identical to the soft function in \eq{S_tau_pT} for $y<0$. 
Our result agrees with ref.~\cite{Procura:2014cba}.\footnote{In the second-to-last expression in eq.~(3.17) of ref.~\cite{Procura:2014cba}, the $\de(k^+ - |\vec k_\perp|)$ term is equal to zero. Due to a typo, the $\pi^2$ term is a factor 2 too big there.} Note that the collinear-soft function for the hemisphere $y>0$  has an identical expression.

\subsection{Two angularities}
\label{sec:tatb}

We now extend \sec{ang} to consider the measurement of two angularities $\tau_a$ and $\tau_b$ as in refs.~\cite{Larkoski:2013paa,Larkoski:2014tva,Procura:2014cba}. Taking $2>b>a$ (and $a,b\neq 1$) implies $\tau_b > \tau_a$ and
\begin{align} \label{eq:f_two_ang}
 \vec f(y,\phi) &= (e^{-|y|(1-a)},e^{-|y|(1-b)}) 
 \,, \quad
 f_\infty(y,\phi)  = e^{-|y|(1-b)}
\,.\end{align}
Writing  $m_{a}=Q \tau^{a}$ and $m_{b}=Q \tau^{b}$, this leads to
\begin{align}
S^\one(\vec m=(m_a,m_b),\mu)
 &=  S^\one(m_b,\mu) 
 \nn \\ & \quad - \theta(m_b - m_a)\,  \frac{2\al_s C_F}{\pi}\! \int_0^{\frac{1}{(b-a)} \ln \frac{m_b}{m_a}} \df y\,  2\Big( \ln \frac{m_b}{m_a} + (a-b) y \Big)
 \nn \\ &=  S^\one(m_b,\mu) 
 - \theta(m_b - m_a)\, \frac{2\al_sC_F}{\pi}\, \frac{1}{b-a} \ln^2 \frac{m_b}{m_a}
\,.\end{align}
This agrees with the expression in ref.~\cite{Larkoski:2014tva}, when converting their angular exponents $\alpha$, $\beta$ to our (current) conventions, $\alpha = 2-a$, $\beta = 2 - b$, and taking into account that they consider only one jet which halves the result.

The corresponding collinear-soft function has again the same amplitude but a modified measurement. For collinear-soft radiation going into the $y<0$ hemisphere,
\begin{align}
 \vec f(y,\phi) &= (e^{y(1-a)},e^{y(1-b)}) 
 \,, \quad
 f_\infty(y,\phi)  = \theta(-y)e^{y(1-b)}+\theta(y)e^{y(1-a)},
\end{align}
which is identical to \eq{f_two_ang} for $y<0$ but not for $y>0$.
This leads to
\begin{align} \label{eq:S1_ma_mb}
{\mathscr S}^\one(\vec m=(m_a, m_b),\mu)  
 &= \frac{1}{2} S^\one(\vec m=(m_a, m_b),\mu) 
 \nn \\ & \quad - \frac{\al_s C_F}{\pi} \int_{0}^\infty\!\df y\, e^{2\eps y(1-a)}  \Big[\frac{1}{\eps} + 2 \ln \frac{\mu}{m_a} + 2\eps \Big(\ln^2 \frac{\mu}{m_a} - \frac{\pi^2}{24}\Big)\Big]  
 \nn \\ & \quad
 - \theta(m_a - m_b) \frac{\al_s C_F}{\pi} \int_{0}^{\frac{1}{(b-a)} \ln \frac{m_a}{m_b}} \!\df y\, 2\Big( \ln \frac{m_a}{m_b} + (a-b) y \Big)
 \nn \\ &=  
 \frac{1}{2} S^\one(\vec m=(m_a,m_b),\mu)  - \frac{1}{2} S^\one(m_a,\mu)
 \nn \\ & \quad
  - \theta(m_a - m_b) \, \frac{\al_sC_F}{\pi}\, \frac{1}{b-a} \ln^2 \frac{m_b}{m_a}
 \nn \\  &=  
 \frac{1}{2} S^\one(m_b,\mu) - \frac{1}{2} S^\one(m_a,\mu)
  - \frac{\al_sC_F}{\pi}\, \frac{1}{b-a} \ln^2 \frac{m_b}{m_a}  
\,. \end{align}
This is consistent with matching the SCET${}_+$ factorization theorem in the bulk with the \SCETa factorization on the boundary, discussed in sec.~4 of ref.~\cite{Procura:2014cba}, since the last term in the second-to-last line of \eq{S1_ma_mb} drops out due to $m_b > m_a$.

It may not be a priori obvious that the collinear-soft function satisfies the kinematic constraint $m_a < m_b$. However,  inserting the collinear-soft scale 
\begin{align}
  \mu_{\mathscr S} = \big(m_a^{b-1} m_b^{1-a}\big)^{1/(b-a)}
\end{align}
in the finite terms gives,
\begin{align}
{\mathscr S}^\one(\vec m=(m_a, m_b),\mu_{\mathscr S})  &=  \frac{\al_s C_F}{\pi} \Big(\frac{1}{b-1}\, \ln^2 \frac{\mu}{m_b} +  
   \frac{1}{1-a}\, \ln^2 \frac{\mu}{m_a} - \frac{1}{b-a} \ln^2 \frac{m_b}{m_a}\Big)
   \nn \\
  &= 0
\,. \end{align}

\section{Conclusions}
\label{sec:conclusions} 

We have presented a convenient method for calculating the effect of soft QCD radiation at one-loop order, for generic $N$-jet processes and measurements. This exploits that soft emissions are uniform in rapidity and azimuthal angle. Through an isolation of the divergent parts, we are able to perform a partial expansion in the regulators already before the integration, simplifying the calculation of the poles and directly leading to an integral for the finite terms. By working with cumulative distributions, complications from plus distributions in intermediate expressions are avoided. As a demonstration of the ease of the calculational framework, soft functions for a range of processes and measurements are computed. We obtain original results for the soft function for $N$-jettiness with generic jet angularities, which required an extension of the hemisphere decomposition~\cite{Jouttenus:2011wh} to make the complicated boundaries between regions tractable, see also ref.~\cite{softfunction}.
We also determine the collinear-soft function for the measurement of two angularities for the first time. 
An automated approach to the two-loop soft function for dijets is underway~\cite{SCETsoft2lTalk}.
Our method reduces the work required for calculating one-loop soft functions, and can for example be applied to calculate the soft functions for the recently introduced XCone class of jet algorithms~\cite{Stewart:2015waa}. 

\begin{acknowledgments}
This work is supported by the European Community under the "Ideas" program QWORK (contract 320389), by the the Netherlands Organization for Scientific Research (NWO) through a VENI grant, and the D-ITP consortium, a program of the NWO that is funded by the Dutch Ministry of Education, Culture and Science (OCW).
\end{acknowledgments}

\appendix

\section{Becher-Bell rapidity regulator}
\label{app:becher}

One may also use the regulator in \cite{Becher:2011dz} to regulate rapidity divergences. This amounts to the substitution 
\begin{align}
\int \df^d k\, \delta(k^2)\, \theta(k^0) \rightarrow \int \df^d k\, \delta(k^2)\, \theta(k^0) \Big(\frac{\nu^-}{k^-} \Big)^\alpha
\end{align}
in the integration over the soft radiation. With $k^-=k_T e^y$ this leads to the replacements 
\begin{align}
|2\sinh y|^\eta  \to  e^{\alpha y}
\,, \quad
\eta \to \alpha
\,, \quad
\nu \to \nu^-
\end{align}
in \eq{start}. With this regulator, \eq{master_eta} gets modified to
\begin{align}
S_{12}^\one(m,\mu)
 &= \frac{\al_s}{2\pi^2}\, \bT_1 \sdt \bT_2\, 
  \int_{-\infty}^\infty\!\df y \int_0^{2\pi}\! \df \phi\, \theta[f(y,\phi)]\, f_\infty(y,\phi)^{2\eps} e^{-\alpha y}
 \\ & \quad \times
 \Big[\frac{1}{\eps} + 2 \ln \frac{\mu\,f(y,\phi)}{m\,f_\infty(y,\phi)} + 2\eps \Big(\ln^2 \frac{\mu}{m} - \frac{\pi^2}{24}\Big)\Big] 
 \Big[1 + \alpha \Big(-\frac{1}{2\eps} + \ln \frac{\nu^-}{m}\Big)\Big]
\,.\nn \end{align}
Note that in order to regulate the integrals, $\alpha$ has to take opposite signs for  $y\to \infty$ and $y\to -\infty$. This is similar to the opposite sign of $\epsilon$ for UV and IR divergencies in dimensional regularization. 

\section{Jet function for transverse thrust}
\label{app:jetfunction}

At one-loop order the jet function contains two emissions. Their contribution to transverse thrust is given by
\begin{align}
  \tau_\perp = \frac{1}{Q \sin^2 \theta} \sum_i \frac{k_{i\,\perp\!\!\top}^2}{2E_i} 
  = \frac{1}{Q \sin^2 \theta} \Big(\frac{k_T^2 \sin^2 \phi}{z Q} + \frac{k_T^2 \sin^2 \phi}{(1-z) Q}\Big) 
  = \frac{s \sin^2 \phi}{Q^2 \sin^2 \theta}
\end{align}
Here $k_{i\,\perp\!\!\top}^2$ is the momentum component perpendicular to the beam \emph{and} thrust axis, $k_T$ is the momentum transverse to the thrust axis (equal and opposite for the two emissions), $\phi$ the azimuthal angle around the thrust axis and $s$ the invariant mass.
Calculating the quark jet function in the approach of ref.~\cite{Ritzmann:2014mka} with no $\eps$-dependence in the $\phi$ integral,
\begin{align}
 J_q^\one(\tau_\perp) &
 = \int_0^\infty\! \df s \int_0^1\,\df z\,\int_0^{2\pi}\,\frac{\df \phi}{2\pi}\,
 \Big(\frac{\mu^2 e^{\ga_E}}{4\pi}\Big)^\eps\, \frac{[z(1-z)s]^{-\eps}}{(4\pi)^{2-\eps} \Ga(1-\eps)}\, \frac{2g^2\,C_F}{s} \Big[ \frac{1+z^2}{1-z} - \eps (1-z) \Big]
 \nn \\ & \quad \times
 \de\Big(\tau_\perp - \frac{s \sin^2 \phi}{Q^2 \sin^2 \theta}\Big)
 \nn \\ &
 = -\frac{\al_s C_F}{\pi}\, \Big(\frac{\mu^2 e^{\gamma_E}}{Q^2 \sin^2 \theta}\Big)^\eps \, \frac{(1-\eps/4) \Ga(1/2+\eps)\,\Ga(2-\eps)}{\sqrt{\pi}\, \eps\, \Ga(1+\eps)\, \Ga(2-2\eps)} \frac{1}{\tau_\perp^{1+\eps}}
\,.\end{align}
Expanding
\begin{align}
 \frac{1}{\tau_\perp^{1+\eps}} = -\frac{1}{\eps}\, \de(\tau_\perp) + \Big(\frac{1}{\tau_\perp}\Big)_+ - \eps \Big(\frac{\ln \tau_\perp}{\tau_\perp}\Big)_+ + \ord{\eps^2}
\,,\end{align}
the finite terms in the one-loop jet function differ from the result in ref.~\cite{Becher:2015gsa} by
\begin{align}
  J_q(\tau_\perp) = J_q^\text{\cite{Becher:2015gsa}}(\tau_\perp) + \frac{\al_s C_F}{\pi}\, \frac{\pi^2}{3}\, \de(\tau_\perp) + \ord{\al_s^2,\eps}
\,.\end{align}

\section{Thrust-like $N$-jettiness}
\label{app:Njettiness}

When all $\alpha_\ell = 0$, the expressions for the soft function contributions given in \sec{1jettiness} simplify and the results of ref.~\cite{Jouttenus:2011wh} are reproduced, as we will show now.
Starting with the hemisphere and the boundary contributions, 
we solve remaining integrals analytically and the sum of \eq{hemii} and \eq{non-hemiA} reduces to the expression~\cite{Jouttenus:2011wh}
\begin{align}
S^\one_{ij,\rm{hemi}}(m_i=\Tau^{i}_{1},\mu)
&= \frac{\al_s}{2\pi}\, \bT_i \sdt \bT_j \Big[ \frac{1}{\eps^2} +\frac{1}{\eps}  \ln \frac{\hat s_{ij} \mu^2}{m_i^2}  + \frac{1}{2} \ln^2 \frac{\hat s_{ij} \mu^2}{m_i^2} - \frac{\pi^2}{12}\Big] \,,
\end{align}
with
\begin{align}
\hat s_{ij} &= \frac{2 q_i' \sdt q_j'}{Q_i Q_j}  =  \frac{4 \omega_i \omega_j}{Q_i Q_j \gamma^2} \, .
\end{align}
Note that for the thrust-like $N$-jettiness the measurement regions are $\phi$ independent and the same result can be obtained without 
the trick of simplifying the hemisphere contributions by splitting off the boundary contributions.

For the non-hemisphere contributions, the integrals in \eq{ints} are simplified by performing the substitutions
\begin{align}
\tilde{y}=\sqrt{\frac{\tilde{n}^0_m-\tilde{n}^3_m}{\tilde{n}^0_m+\tilde{n}^3_m}} e^y\,, \quad \tilde{\phi} = \phi - \arccos \sqrt{\frac{(\tilde{n}^1_m)^2}{(\tilde{n}^1_m)^2+(\tilde{n}^2_m)^2}}\,,
\end{align} 
which leads to
\begin{align}
 \tilde{I}_0 (\alpha_\ell=0) &=\frac{1}{\pi} \int_{0}^\infty\! \frac{\df \tilde{y}}{\tilde{y}} \int_{-\pi}^{\pi}\! \df \tilde{\phi}\, \theta \Big( \tilde{y}^2 - \frac{\hat s_{im}}{\hat s_{jm}}\Big)\, \theta \Big( \frac{\hat s_{ij}}{\hat s_{jm}} -1 -\tilde{y}^2 + 2 \tilde{y} \cos \tilde{\phi} \Big) \,,
 \nn \\
 \tilde{I}_1 (\alpha_\ell=0) &= \tilde{I}_0  (\alpha_l=0) \ln \Big( \frac{\hat s_{jm}}{\hat s_{ij}} \Big)  +  \frac{1}{\pi} \int_{0}^\infty\! \frac{\df \tilde{y}}{\tilde{y}} \int_{-\pi}^{\pi}\! \df \tilde{\phi}\, 
 \nn \\ & \quad \times
 \theta \Big( \tilde{y}^2 - \frac{\hat s_{im}}{\hat s_{jm}}\Big) \theta \Big( \frac{\hat s_{ij}}{\hat s_{jm}} -1 -\tilde{y}^2 + 2 \tilde{y} \cos \tilde{\phi} \Big)
 \ln (\tilde{y}^2 +1 - 2 \tilde{y} \cos \tilde{\phi}) \,,
\end{align} 
with
\begin{align}
\hat s_{im} &= \frac{2q_i' \cdot q_m'}{Q_i Q_m}  = \frac{2 \omega_i \omega_m}{Q_i Q_m \gamma} (\tilde{n}^0_m-\tilde{n}^3_m)
\,, \quad
\hat s_{jm} = \frac{2q_j' \cdot q_m'}{Q_j Q_m} = \frac{2 \omega_j \omega_m}{Q_j Q_m \gamma} (\tilde{n}^0_m+\tilde{n}^3_m)\,.
\end{align}
This is in agreement with the non-hemisphere expression of ref.~\cite{Jouttenus:2011wh}. There 
the remaining integrals have been further simplified to one-dimensional integrals.

\bibliographystyle{jhep}
\bibliography{refs}

\end{document}